\documentclass[amssymb,amsmath,showpacs,superscriptaddress,preprint]{revtex4-1}
\usepackage{graphicx}
\usepackage{amsmath}
\usepackage{xcolor}
\begin{document}
\title{Cycle deformation enabled controllable mechanical polarity of bulk metallic glasses}
\author{Bao-Shuang Shang}
\affiliation{Songshan Lake Materials Laboratory, Dongguan 523808, China}
\affiliation{Beijing Computational Science research center, Beijing 100094, China}
\author{Wei-Hua Wang}
\affiliation{Songshan Lake Materials Laboratory, Dongguan 523808, China}
\affiliation{Institute of Physics, Chinese Academy of Sciences, Beijing 100190, China}
\author{Peng-Fei Guan}
\email{pguan@csrc.ac.cn}
\affiliation{Beijing Computational Science research center, Beijing 100094, China}
\date{\today}
\begin{abstract}
Tuning anisotropy in bulk metallic glasses, ideally isotropic, is of practical interest in optimizing properties and of fundamental interest in understanding the amorphous structure and its instability. 
By employing {the quasi-elastic} asymmetric mechanical cycling method, we {effectively} induce the mechanical polarity of a model bulk metallic glass, without damaging the sample or introducing significant annealing or rejuvenation effects. 
Moreover, the polarized anelastic limit can be well controlled by regulating the amplitude of mechanical cycling.
Through the atomic-level analysis of nonaffine displacement, we find that only plastic atomic rearranged events corresponding to the training direction can be exhausted by {asymmetric cycling} and the survived anelastic events dominate the directional anelastic limit.
The polarized distribution of local yield stress reveals that the mechanical polarity is attributed to the {plastic-event-healing} induced asymmetry of local potential energy surface, rather than frozen-in anelastic strain. 
Furthermore, the {healing} of plastic events associated with the redistribution of local residual stress indicates the origin of polarity induced by asymmetric cycling.
Our study is of fundamental importance, which further our understanding of the mechanical deformation of metallic glasses and shed some light on the prospects for improved properties through induced anisotropy.
\end{abstract}
\maketitle
\section{introduction}
As-cast bulk metallic glasses (BMGs) are commonly expected to be isotropic \cite{Greer200914}, which are freshly formed from a glass-forming liquid composed of atoms that are not orientable \cite{Cheng2011379}. 
Thus, the atomic structure and related mechanical responses of as-quenched BMGs are isotropic up to small fluctuations. {It means that the stress-strain curve under an external strain should exhibit symmetry for positive or negative loading directions, in general, which can be well described by isotropic mechanics\cite{Wang2012a}.}
However, recent studies suggest that anisotropy can be induced extensively by thermomechanical treatments, such as homogeneous mechanical creep \cite{Dmowski2010Structural,Concustell2011Induced,PhysRevB.35.2162,PhysRevB.48.3048,PhysRevLett.105.205502}, homogeneous uniaxial tension or compression \cite{Ott2008Anelastic,Sun2016}, torsion \cite{sun2014pure}  and cyclic loading \cite{LouzguineLuzgin2017}. 
{The typical mechanical anisotropy in amorphous solid is sometimes referred as the Bauschinger effect \cite{sun2014pure}, and the corresponding structural origin should be different from anisotropic elasticity in crystalline alloys associated with dislocations and microstructural features.} Thus, it has long been of interest to understand the origin of anisotropy in MGs \cite{PhysRevB.35.2162,PhysRevB.48.3048}, and thereby the mechanisms of flow and plasticity.

Though mechanical anisotropy in bulk metallic glasses has been studied broadly in both experiments and simulations, however, published results have mostly observed anisotropy or polarity by thermomechanical treatments under high strain plastic flow deformation \cite{Ott2008Anelastic,PhysRevE.82.026104} or high temperature creep \cite{Dmowski2010Structural}.
It is well accepted that frozen-in anelastic strain \cite{Egami2013} in the MGs is associated with, and indeed must lead to, anisotropy in structure and mechanical properties. 
Recent numerical study \cite{PhysRevLett.124.205503} suggests that the structural origin of the Bauschinger effect in steady state flow of a two-dimensional binary Lennard-Jones model glass is the polarization of local residual strength. 
{All these studies indicate that mechanical anisotropy always stores in the plastically-deformed MG samples.
It’s hard to induce mechanical anisotropy without destroying metallic glass sample based on existing methods.} Thus, it is of interest to achieve controllable anisotropy or polarity of MGs without severe plastic deformations.

According to the ``shear transformation zone (STZ)'' theory \cite{PhysRevE.57.7192}, in amorphous systems, the triggering of local plastic events is sensitive to the loading direction. 
It is conjectured that the densities of STZs for positive and negative strains, denoted as $n_+$ and $n_-$ could be distinct. 
Thus, if we can regulate the normalized difference between these, in principle, mechanical anisotropy or polarity can be successfully induced into the MGs.
Early studies \cite{sinning1985irreversible,friedrichs1989study}  of melt-spun ribbons and thin films show that soft-magnetic metallic glasses can be significantly anisotropic by thermal annealing. 
Thermal effects are usually isotropic and the thermal-annealing induced anisotropy in ribbons or films can be attributed to intrinsic structural anisotropy and symmetry-breaking nature of surface mobility. 
Bulk metallic glasses (BMGs) are expected to be much closer to isotropy in their as-cast state and the thermal annealing should not be an effective way to produce significant anisotropy. 
The stress-temperature scaling \cite{PhysRevLett.104.205701}  has quantified the intimate coupling of temperature and shear stress in inducing local atomic rearrangements and thereby liquid-like flow in glasses.
It is confirmed by previous works \cite{sun2016thermomechanical} that both thermal and mechanical treatments can regulate the densities of STZs and thereby the physical properties of MGs.
Comparing with thermal stimulus, the influences of mechanical loading are directional and anisotropic \cite{greer2016stored}.
Based on the symmetrical mechanical cycling protocol with various strains at finite temperature, the memory \cite{PhysRevLett.112.025702}, ageing \cite{das2018annealing,PhysRevLett.124.225502} or rejuvenation \cite{leishangthem2017yielding}  effect can be successfully induced in bulk amorphous solids, respectively.
These symmetry training protocols should bring the similar influences to the densities of STZs for positive and negative strains, which cannot tune the sample polarity as expected.
The present work explores whether the normalized difference between the densities of two kinds of STZs, associating with polarity in BMGs, can be regulated by breaking the symmetry of deformation amplitude of mechanical cycling.
This may be of practical interest in optimizing properties by accessing new glassy states, and of fundamental interest in probing the precise nature of the STZs and the mechanisms of plasticity.

In this work, we successfully induce the mechanical polarity into a model bulk metallic glass by employing an asymmetric mechanical cycling (AMC) method. 
{As the cylcling amplitude is in the quasi-elastic region, the trained sample retains energy state and shape, however, a polarized plastic response can be obersved. Moreover, the polarized anelastic limit can be well regulated by the cycling amplitude.}
Using the local yield stress as an indicator, we prove that this mechanical polarity is inherently come down to the non-symmetry topography of potential energy landscape (PEL), which is ascribed to the annihilation of pre-existing sub-basins. 
{The close link between the atomic rearrangements and the redistribution of local residual stress indicates that the polarity induced by the AMC originates from the enhausion of plastic rearrangements, accosiated with the survival of anelastic rearrangements, in the training direction.}
Our finding provides a direct evidence that the anisotropy can also be achieved by regulating the local plasticity, resulting from decoupling among the plastic STZs for positive and negetaive strains in metallic glasses.

\section{Simulation method}
\subsection{As-cast sample preparation}

The molecular dynamic simulations were conducted by LAMMPS \cite{plimpton1995fast} with embedded atom method (EAM) potential \cite{mendelev2009development}, a typical model metallic glass system $Cu_{50}Zr_{50}$ was investigated which contained 8 000 atoms. 
The samples firstly were kept in equilibrium at 2000 K for 1 ns, then quenched to 1 K with quench rate $10^{13} \text{ K/s}$. The pressure and temperature were controlled by isotherm-isobaric thermostat (NPT) with Nose-Hoover thermostat \cite{nose1984unified}, and the environmental pressure was maintained at 0 bar. 
The as-cast samples in shear strain-free state were obtained by minimizing the as-quenched configurations at 1 K. Ten independent samples were prepared to acquire statistical data.

\subsection{Asymmetric mechanical cycling (AMC) protocol}

To elude temperature influences and utral-high strain rate effects in simulations, 
the cyclic deformation was carried out in simple shear, using the athermal quasistatic protocal \cite{rodneyandtanguy2011}, in which the system was deformed by a small increment of affine strain ($\Delta \gamma_{xy}=10^{-4}$) followed by energy minimization. 
In order to induce asymmetric influence, we broke the symmetry of deformation amplitude of mechanical cycling.
As shown in Fig. \ref{fig:1}(a), we first loaded the as-cast sample at shear strain-free state (`O') along the selected training direction (denoted as ``$+$'') to the deformed state `A' with training strain amplitude $\gamma_A$. Then, the deformed sample was unloaded in the opposite direction (denoted as ``$-$'') to the shear strain-free state `B' with $\gamma_{xy} =0$. 
To release macro residual stress, the sample was relaxed to shear stress-free state `C' with $\sigma_{xy} =0$. 
By repeating the subsequent cycling process (C-A-B-C), all physical properties of the cycled sample were measured.
This is {a} strategy to tune effectively strain control with relaxation process to stress control process, whereas stress control process can not access beyond the yield strain, and it can avoid the notable stress fluctuation.
In the quasi-elastic regime, those two processes are equivalent. 
The different strain amplitudes,  $\gamma_A= 0.05$, 0.06 or 0.08, were employed to prepare the trained sample with different mechanical cycling histories.
The potential energies of the shear stress-free states are monitored as a function of cycle number $N$ under various strain amplitudes.  

\subsection{Physical property characterization}

\subsubsection{{Radical distribution function (RDF) and related anisotropy indicators}}

{To characterize atomic structural polarization, the angular dependent radical distribution function (RDF) $g_{\theta}(r)$ was defined   \cite{Ingebrigtsen2017sturcture} as :
\begin{equation*}
g_{\theta}(r)=\frac{1}{2r^2\rho N} \sum_{i \neq j} \delta(r-|r_{ij}|) \delta(\theta-\theta_{ij})
\end{equation*}
Where $\rho$ was number density, $N$ was atom number. And the angular dependent RDF function can be expanded by the spherical harmonic function $Y_l^{m}(\theta)$ as :
\begin{equation*}
  g_{\theta}(r)=\sum_{l,m} g_{l}^{m}(r) Y_{l}^{m}(\theta)
\end{equation*}
For axial symmetry, m equals 0, and odd term equals 0. The anisotropic terms with $l>2$ can be neglected. For $m=0$ and $l=2$, the $g_2^{0}(r)$ term represents the anisotropy part of RDF.}

The angular dependent two-body excess entropy $s_{2}^{\theta}$ was defined as
\begin{equation*}
s_{2}^{\theta}=-\rho \int_{0}^{\infty} [g(r,\theta)\ln g(r,\theta)-(g(r,\theta)-1)]r^2dr
\end{equation*}

\subsubsection{Local yield stress ($\tau_{y}$) and local residual stress ($\tau_0$)}

According to the method developed by Patinet et al \cite{PhysRevLett.117.045501}, the local yield stress was calculated as shown in Fig. \ref{fig:S1} in Supplementary Material (SM), {the sample was first meshed into $40 \times 40 \times 40$ grids, and the local yield stress at each grid point was calculated by investigating a spherical region, centered at the grid point, with a radius of 1 nm. 
Each spherical region contains about 250 atoms, which is comparable with the average size of STZs (tens to hundreds atoms) and still holden by the linear consititutive relation.}
For each grid point, the whole sample was sheared with affine deformation and only the investigated region was relaxed.
The outside region went through affine deformation,the stress-strain curve of the local region was recorded, and the first stress drop was termed as local yield stress (LYS). 
The initial stress of each grid point was termed as local residual stress.

\subsection{Local atomic rearrangement}

To identify atomic rearrangements during the AMC process,
we used $D_{min}^{2}$  \cite{PhysRevE.57.7192} which was calculated as 
\begin{eqnarray*}
D_{j}^{2} &=& \frac{1}{N_{j}}\sum_{k}[\mathbf{r}^{tar}_{jk}-\mathbf{\gamma}_{j} \cdot {\mathbf{r}^{ref}_{jk}}]^2 \\
\mathbf{\gamma}_{j} & \equiv& \mathbf{Y}_{j}^{-1}\cdot \mathbf{X}_{j} \\
\mathbf{Y}_{j} &=& \sum_{k} [({\mathbf{r}^{tar}_{jk}})^T \cdot {\mathbf{r}^{ref}_{jk}}] \\
\mathbf{X}_{j} &=& \sum_{k} [({\mathbf{r}^{tar}_{jk}})^T \cdot {\mathbf{r}^{ref}_{jk}}]
\end{eqnarray*}

Where ${\mathbf{r}}_{jk}^{ref} \equiv {\mathbf{r}}_{k}^{i}-{\mathbf{r}}_{j}^{ref}$, ${\mathbf{r}}_{k}^{ref}$ was the radius vector of particle $k$ in reference configuration and ${\mathbf{r}}^{tar}_{jk}$ was the vector from particle $k$ to particle $j$ in target configuration. 
The atoms within the top 5\% value of $D_\text{min}^2$ are identified as rearrangemented atoms. The selection of the threshold value (3\%-14\%, Fig. \ref{fig:S5}) does not affect the conclusions.
The anelastic rearrangements in a sample, which are reversible upon unloading, can be distinguished by calculating the $D_\text{min}^2$ between the deformed and the unloaded configurations, respectively. The plastic rearrangements in the sample, which are irreversible relative to the initial state, can be distinguished by calculating the $D_\text{min}^2$ between the initial and the unloaded configurations. 

\section{Results}
\subsection{Potential energy and mechanical hysteresis loop evolution during mechanical cycling}

\begin{figure}[!htpb]
\centering
\includegraphics[width=0.8\textwidth]{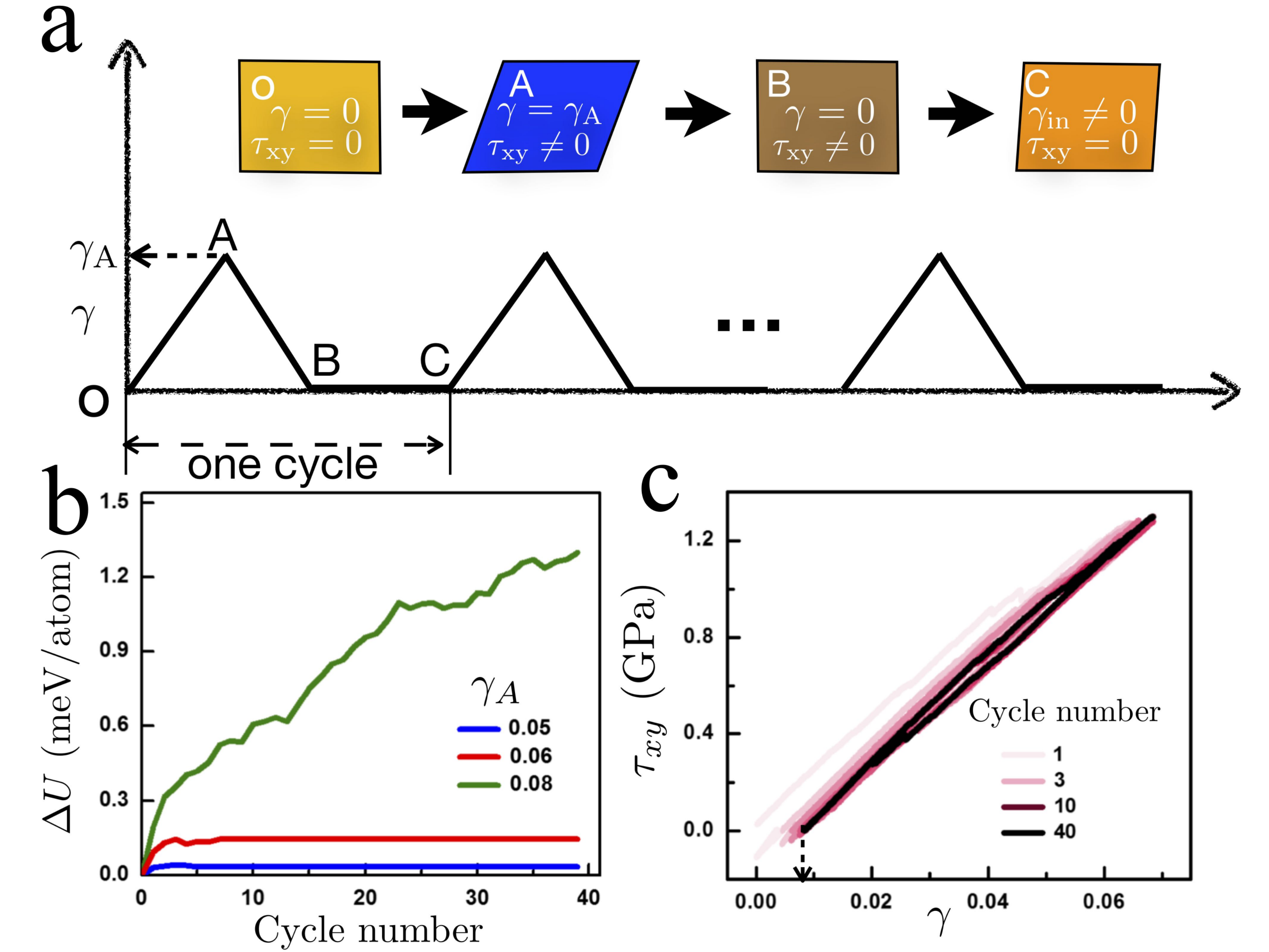}
\caption{Procedure of asymmetric mechanical cycling (AMC): (a) Illustration of the AMC process: firstly, the sample is sheared to $\gamma_{max}$ which is termed as training direction (\textbf{O} $\to$ \textbf{A}), and unloading back to the original geometry (\textbf{A} $\to$ \textbf{B}), then release the macro residual stress to zero (\textbf{B} $\to$ \textbf{C}), repeat the above process up to certain cycle number $N$.
Stress and strain along the training direction ($+$) is positive, and along the opposite direction ($-$) is negative.
(b) atomic potential energy evolves with cycle number for different training amplitude, $\gamma_A=0.05$, $\gamma_A=0.06$ and $\gamma_A=0.08$, respectively. The macro yield strain for the as-cast sample is 0.1.(c) The simulated strain-stress curves during the AMC with $\gamma_A=0.06$. The arrow marks the residual plastic strain ($\sim 0.7 \%$) of the trained sample after 40 cycles with respect to the as-cast sample.}
\label{fig:1}
\end{figure}

The potential energy of the shear stress-free states of the cycled samples under various training amplitudes $\gamma_A$, with respect to the as-cast state, as a function of the cycle numbers $N$ are shown in Fig. \ref{fig:1} (b).
{As the training amplitude amplitude $\gamma_A=0.08$, approaching to the macro yield strain ($\gamma_y \sim 0.1$), the potential energy increases monotonously with the cycle number $N$ and significant rejuvenation can be achieved by the AMC. It is qualitatively similar with symmetric oscillatory shear such as Ref \cite{PhysRevE.88.062401,PhysRevE.88.020301,PhysRevE.89.062307}, and there would be a yield transition lower than macro yield strain \cite{leishangthem2017yielding}, due to the accumulation of atomic rearrangements.
However, for an amplitude far below the yield strain, such as $\gamma_A=0.05$ or 0.06, the potential energy first increases gently and reaches a platform after finite cycles. It implies that the energy state of cycled sample can quickly achieves a saturation state with light rejuvenation, as the training amplitude $\gamma_A$ belongs to the quasi-elastic regime.
In this paper, we typically show the results for training amplitude $\gamma_A=0.06$. The results for $\gamma_A <0.06$ are qualitatively similar, except the anelastic limit at the training direction.} More details will be discussed in the discussion section. The simulated strain-stress curves during the AMC with $\gamma_A=0.06$ are shown in Fig. \ref{fig:1}(c). 
Hysteresis characteristic can be observed for each loop due to the dissipation of mechanical energy during cycling loading which is an intrinsic behavior of amorphous solids.
The area contained in the hysteresis loop decreases towards a constant value with the cycle number $N$ increasing.
It indicates that the energy dissipation originated from the number of activated atomic rearrangements decreases 
first and reaches saturation during the AMC.
Moreover, the unclosed loop for one loading-unloading cycle implies irreversible of some atomic rearrangements during 
loading with respect to the undeformed state, whereas a closed loop indicates the completely reversible of atomic rearrangements.
{As the cycle number $N$ increases, the loop tends to be close which suggests the exhaustion of irreversible atomic rearrangements.
All rearrangements in the sample with high cycle number are reversible.
As a result, the final shape of the shear stress-free sample does not change with $N$ and a constant residual plastic strain ($\sim 0.7 \%$) with respect to the as-cast sample can be observed.
The effective regulation between the plastic and anelastic rearrangements by the AMC protocol may offer us the opportunity for achieving controllable anisotropy or polarity of MGs without huge plastic deformations.}

\subsection{Mechanical polarity and local non-affinity of trained sample}

\begin{figure}[!htpb]
\centering
\includegraphics[width=0.8\textwidth]{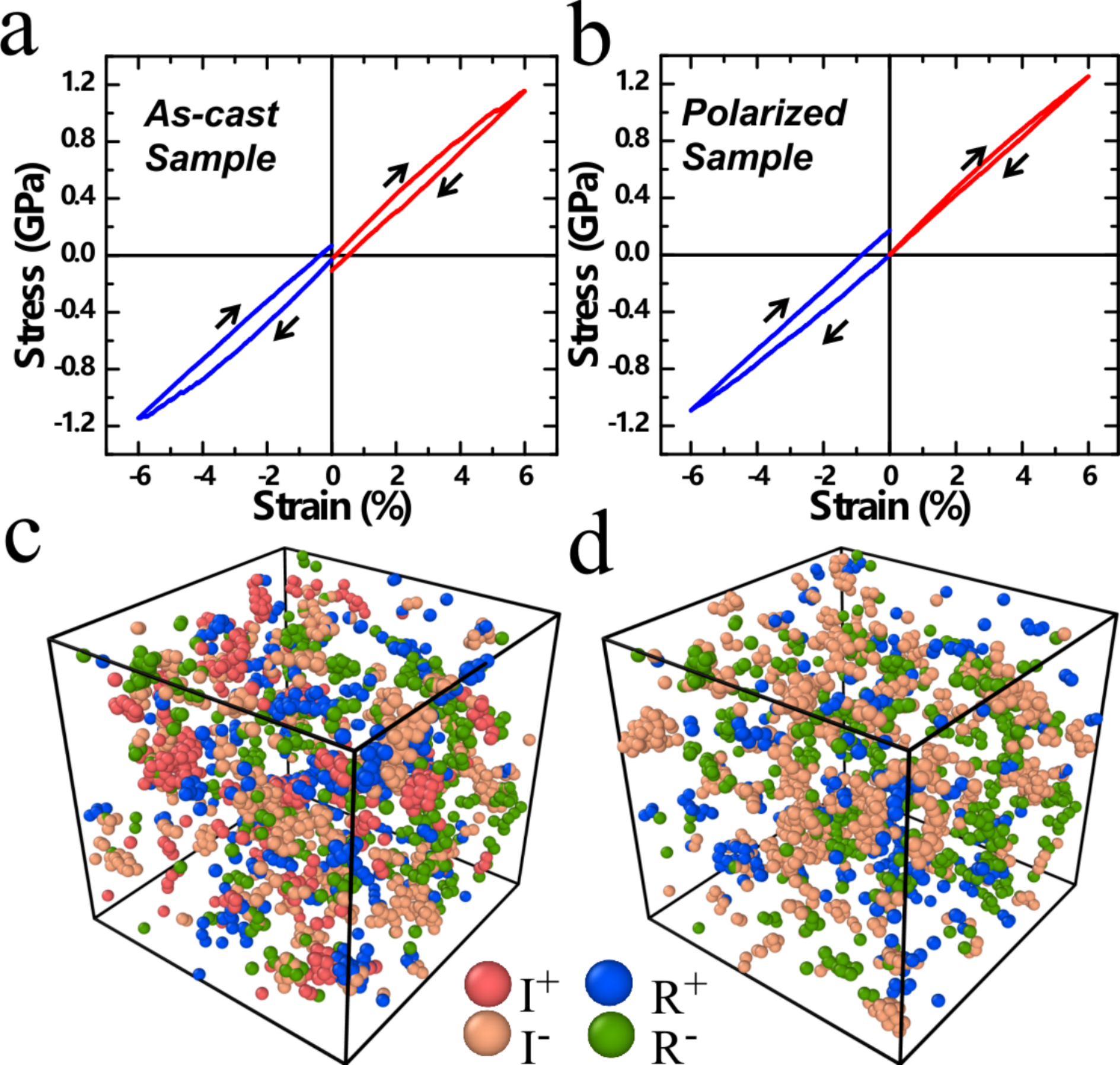}
\caption{Mechanical properties of as-cast and polarized samples in two opposite directions under the mechancial loading-unloading test with  $\gamma_{max}=0.06$. 
The cyclic stress-strain curves of as-cast (a) and trained (b) samples in two opposite directions: positive ($+$) and negative ($-$), respectively.
Atomic rearrangements in as-cast (c) and polarized (d) samples during loading-unloading cycles in positive and negative directions.
$I^+$, $I^-$ represent plastic atomic rearrangements in positive and negative directions, and $R^+$, $R^-$ represent anelastic atomic rearrangements in positive and negative directions, respectively.}
\label{fig:2}
\end{figure}

After 40 cycles,  the samples at shear stress-free state are selected as the trained samples. 
They behave loading directional dependent stress-strain curves (Fig. \ref{fig:S2} in the Supplementary Material (SM)).
To highlight this polarity behavior, the mechanical loading-unloading tests ($\gamma_A=0.06$) were performed at two opposite directions: marked as positive ($+$) for the training direction and negative ($-$) for the opposite direction.
The representative measured cyclic stress-strain curves of as-cast and trained samples are listed in Fig. \ref{fig:2}(a) and (b), respectively.
For the as-cast sample, whatever it's the positive direction or the negative direction, the system can not return to its initial state after the loading-unloading process.
Residual stress or strain can be observed in the shear strain or stress-free unloaded state for both two directions (Fig. \ref{fig:2} (a)).
The as-cast MG sample is mechanical isotropy, which is consistent with previous results. However, for the trained sample, the cyclic stress-strain curves are distinct in two directions (Fig. \ref{fig:2} (b)).
The positive-direction loading-unloading process can bring the system completely back to its initial state, whereas the negative direction process can not.
It confirms the mechanical polarity or anisotropy in trained MG sample and offers us the opportunity for understanding the atomic level mechanism of the anisotropy induced by the AMC protocol.

   The anisotropic  mechanical hysteresis loops indicate distinct dissipation behaviors in two opposite directions of trained sample.
To reveal the atomic level responses related to the energy dissipation, the atomic rearrangements during loading-unloading cycle for both positive and negative directions are characterized in as-cast and trained samples.
As we defined in the Method section II(D), the atomic rearrangements can be distinguished as irreversible ( denoted as `I' ) and reversible ( denoted as `R') rearrangements. As shown in Fig. \ref{fig:2} (c) and (d), the triggering of atomic rearrangements is sensitive to the loading direction.
For the as-cast sample (Fig. \ref{fig:2} (c) ), though the spatial distributions of activated rearrangements for positive ($I^+$, $R^+$) and negative ($I^-$, $R^-$) directions are distinct, the densities of positive and negative rearrangements are almost identical.
Thus, the mechanical properties of the as-cast sample should be isotropic (Fig. \ref{fig:2} (a) ) according to the STZ theory \cite{PhysRevE.57.7192}.
However, as shown in Fig. \ref{fig:2} (d), the densities of atomic rearrangements for positive and negative directions are distinct in trained sample, particularly the irreversible (plastic) rearrangements ($I^+$ and $I^-$).
The density of $I^+$ equals to 0 in the trained sample, which  means the mechanical behaviors under positive strain is dominated by the existed reversible (anelastic, $R^+$) rearrangements.
As a result, the trained sample presents a closed loop for positive direction cycle (Fig. \ref{fig:2} (b)).
The nonzero density of $I^-$ confirms the unclosed mechanical hysteresis loop for negative direction.
These provide the direct evidences that mechanical anisotropy or polarity has been successfully induced into the MGs by regulating the normalized difference between the plastic rearrangements ($I^{+}$ and $I^{-}$), as conjectured by STZ theory \cite{PhysRevE.57.7192}.
Moreover, only the plastic rearrangements in the training direction can be regulated by the AMC. 
{It is obviously different with the thermal annealing induced STZ reduction associated with the notable potential energy decrease \cite{PhysRevE.88.062401}.
Instead, the potential energy of these rearrangements can be regarded as the ``plastic-events-healing'' during the AMC.
The AMC trained sample provides an extreme MG example that the plastic STZs for positive and negative strains are full decoupled.
It is of particular interest to investigate the structural origins of this spacial mechanical polarity induced by the AMC.}
\subsection{Structural indicator of the mechanical polarity}

\begin{figure}[!htpb]
\centering
\includegraphics[width=0.8\textwidth]{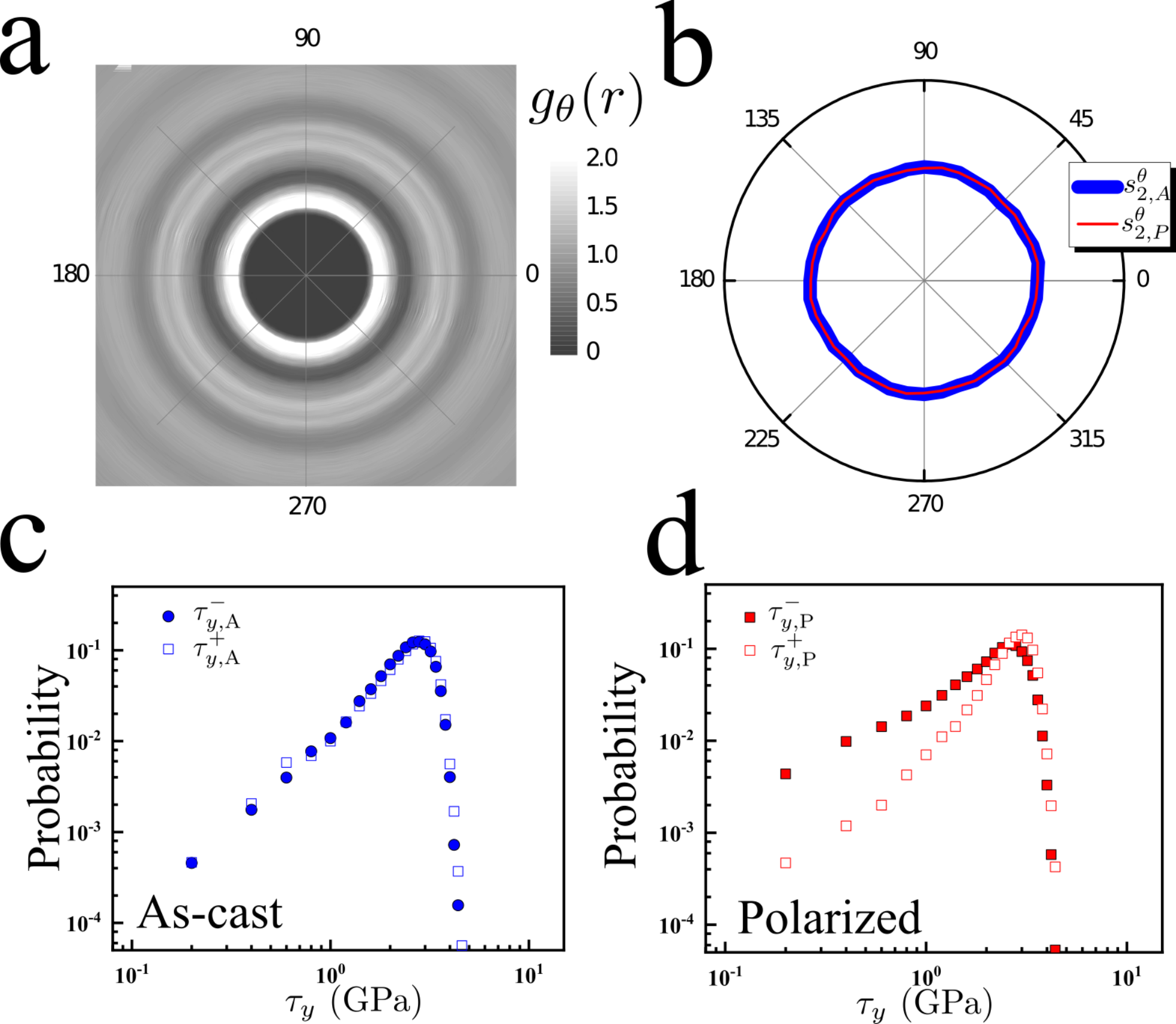}
\caption{{Anisotropy structural information in polarized sample. (a) The angular-dependent radical distribution function (RDF) $g_{\theta}(r)$ of polarized sample. (b) The angular-dependent excess entropy of as-cast ($S_{2,A}^{\theta}$) and polarized sample ($S_{2,P}^{\theta}$), respectively. (c), (d) Probability distribution of LYS for as-cast ($\tau_{y,A}^{+},\tau_{y,A}^{-}$) and polarized sample ($\tau_{y,P}^{+},\tau_{y,P}^{-}$), respectively.}}
\label{fig:3}
\end{figure}

Fig. \ref{fig:3}(a) shows the angle-dependent radial distribution function (RDF) of the polarized sample, in odds to the loading state observed in the reference \cite{luo2015mechaincal}, there is no directional preference in the global structure level. 
Furthermore, in contrast with plastic flow regime \cite{Ingebrigtsen2017sturcture}, there is also no significant difference (Fig. \ref{fig:3}(b)) between the excess entropies of as-cast and polarized samples.
The calculated anisotropic part of RDFs ($g_{2}^{0}(r)$) for as-cast and polarized samples (Fig. \ref{fig:S3}) are also similar.
{
Contrast with the significant anisotropy in plastic flow state \cite{peng2013stress}, the polarized sample doesn't show any signal of anisotropy in its RDF and related indicators.}

{Since the local yield stress (LYS) \cite{PhysRevB.35.2162,PhysRevB.48.3048,PhysRevLett.105.205502} has been proved as a good indicator to predict the loading orientation sensitive of local atomic rearrangements, the ``plastic-events-healing'' in trained direction induced mechanical anisotropy may also strong correlated with the orientation-dependent LYS behaviors.
According to the definition of the LYS, this indicator is intrinsically determined by the local atomic configuration and cab be regarded as a local structural indicator of MGs.}
The probability distributions of the calculated LYS for negative and positive directions of as-cast and polarized samples are shown in Fig. \ref{fig:3} (c) and (d), respectively.
The as-cast sample presents almost identical distributions (Fig. \ref{fig:3}(c) and Fig. \ref{fig:S4}(a)) for negative and positive directions, which is consistent with previous works.
However, the probability distribution of LYS for the polarized sample shows a notable discrepancy between negative and positive directions (Fig. \ref{fig:3}(d) and Fig. \ref{fig:S4}(b)), especially, at low value regime

The value of LYS reflects the required local applied stress to trigger rearrangement in corresponding local region, and the lower LYS value means the region prone to rearrangement at lower applied stress.
The notable discrepancy between negative and positive directions indicates that much more regions could be triggered at low stress under negative direction loading in the trained sample.
It arises the distinct mechanical responses in two opposite directions intrinsically.
Therefore, the orientation-dependent probability distribution is strongly correlated with mechanical deformation behaviors and the LYS can be regarded as the structural indicator of the mechanical polarity induced by the AMC.
{The asymmetry distribution of LYS between negative and positive directions has also been observed in plastic flow state \cite{PhysRevLett.102.235503,PhysRevLett.124.205503}. 
However, contrast to the memory lost of atomic structure during the plastic flow deformation, the polarized samples in this work do not undergo a global structure relaxation. Thus, it is of interested to investigate the link between the local atomic rearrangements and the polarized LYS evolution during the AMC with $\gamma_A<0.08$.
}
\subsection{Local structural evolution during the AMC}

\begin{figure}[!htpb]
\centering
\includegraphics[width=0.8\textwidth]{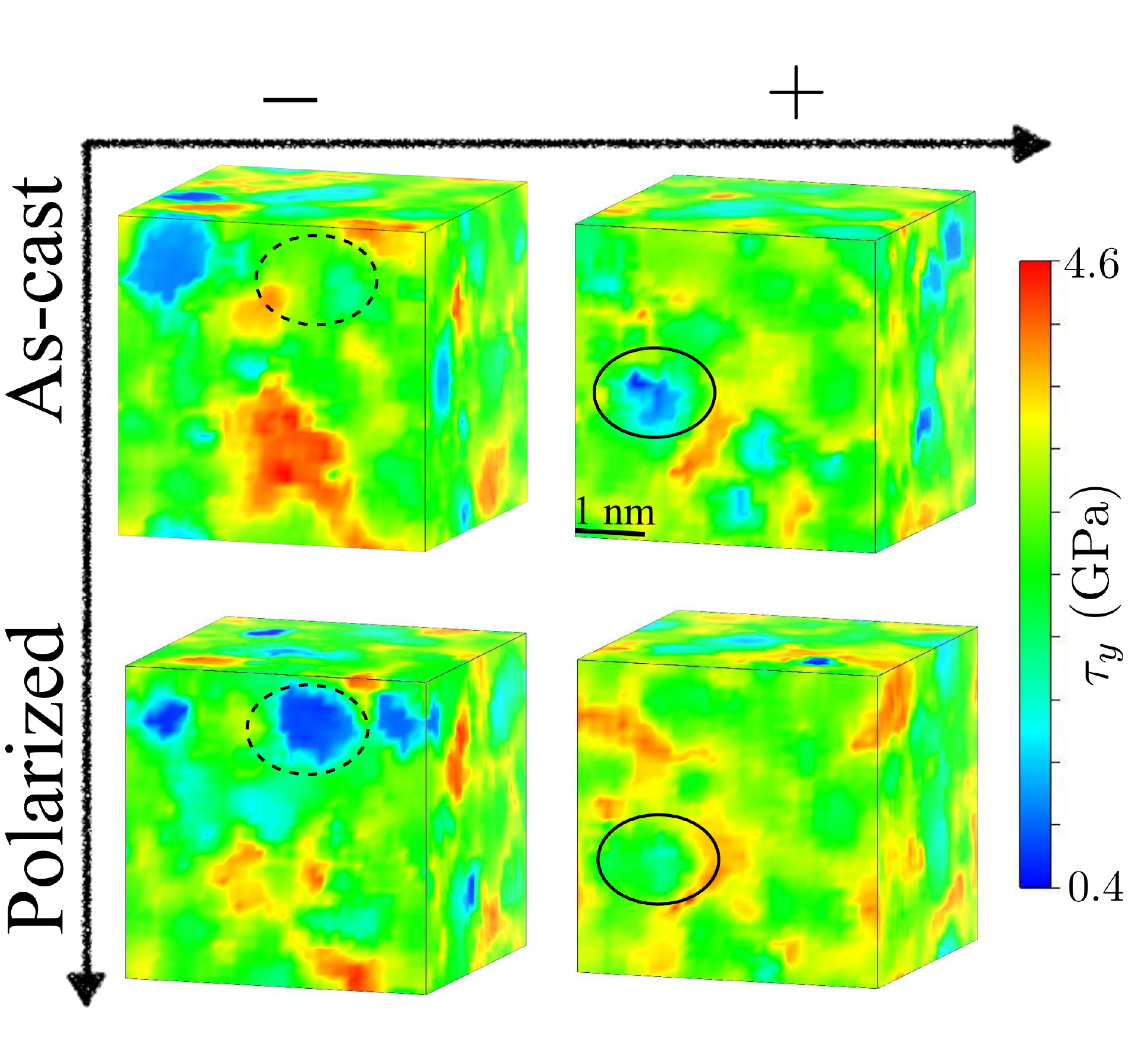}
\caption{{3D view of local yield stress (LYS) for as-cast and polarized sample, respectively. From left to right are the spatial distribution in negative and positive directions, respectively. From top to bottom are the spatial distribution of as-cast and polarized sample, respectively.   Solid circles mark the representative hardening zone, and dotted circles mark the representative softening zone. }}
\label{fig:4}
\end{figure}

\begin{figure}[!htpb]
\centering
\includegraphics[width=0.8\textwidth]{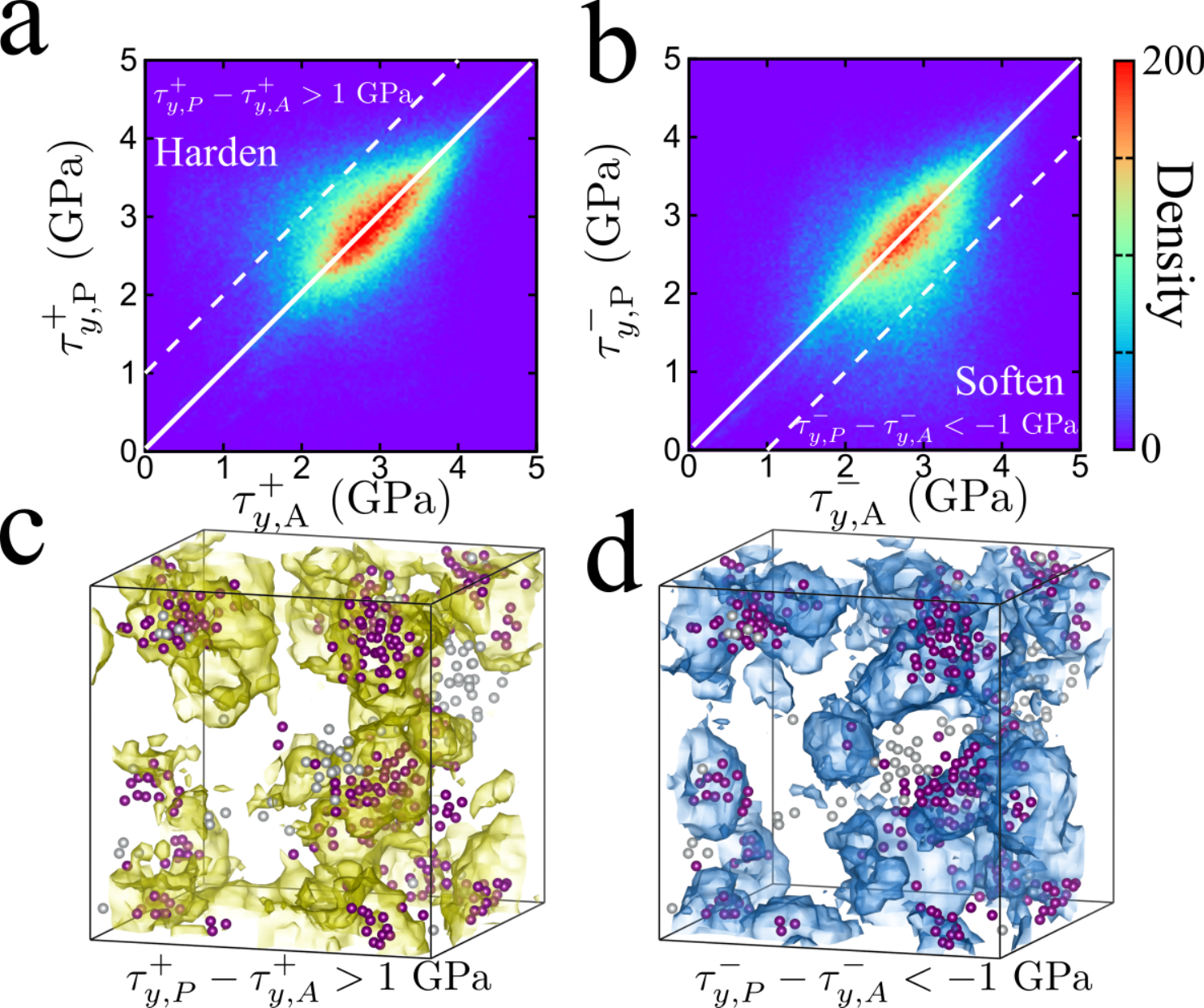}
\caption{{Structure information of the polarized zone.
(a),(b) density contour plot between as-cast and polarized configuration, $\tau_{y,A}^{+}$ vs $\tau_{y,P}^{+}$ is the LYS of positive direction for as-cast and polarized configuration, respectively, $\tau_{y,A}^{-}$ vs $\tau_{y,P}^{-}$ is the LYS of negative direction. 
The dash line is guided for the eyes shows $\tau_{y,P}^+-\tau_{y,A}^{+} =1$ GPa and $\tau_{y,P}^--\tau_{y,A}^{-}=-1$ GPa, respectively. And the solid line shows the unchanged line $\tau_{y,P}-\tau_{y,A}=0$ for positive and negative direction. (c), (d) the 3D view of the polarized zone for $\tau_{y,P}^{+}-\tau_{y,A}^{+} >1$ GPa, and $\tau_{y,P}^{-}-\tau_{y,A}^{-} <-1$ GPa, respectively. The purple atom is the plastic rearrangement atom during the AMC \emph{involved} in the polarized zone, and the gray atom is the plastic rearrangement atom \emph{outside} the polarized zone.}}
\label{fig:5}
\end{figure}

{The spatial distributions of LYS in negative and positive directions are shown in Fig. \ref{fig:4}. No matter for as-cast sample or for polarized sample, the LYS presents inhomogeneous distribution which is the intrinsic behavior of amorphous materials.
On the other hand, there is no strong coupling between the negative and positive LYS distributions ( Figs. \ref{fig:S4} ), which implies the protocol-dependent local rearrangements in MGs.
More important, with respect to the as-cast sample, softening zones can be obviously observed in the negative direction, whereas hardening zones were popular in the positive direction for the polarized sample.
The distinct evolutionary trends in two opposite directions determine the mechanical polarity of training sample. 
Figs. \ref{fig:5} (a) and (b) show the spatial correlations of the LYS between as-cast and polarized configurations in two opposite directions.
The solid line illustrates the perfect correlation as the LYS distribution is not changed.
The correlation of positive LYS deviating to the top left suggests the hardening in positive direction, whereas the correlation of negative LYS deviating to the bottom right implies the softening in negative direction of the polarized sample.
It confirms the results in Figs. \ref{fig:3} and \ref{fig:4}.
The hardening and softening zones, served as the polarized zones, should play the key role in the AMC-induced mechanical polarity.
To characterize these zones,in the polarized sample with respect to the as-cast sample, we choose 1 GPa as threshold value (dash line in Fig. \ref{fig:5} (a), (b)), and the polarized zones are distinguished by the corresponding $\Delta \tau_{y}^{+}=\tau_{y,P}^{+}-\tau_{y,A}^{+}> 1$ GPa (hardening zone) or $\Delta \tau_{y}^{-}=\tau_{y,P}^{-}-\tau_{y,A}^{-} <-1$ GPa (softening zone).
The major conclusions are insensitive to the threshold value (1GPa, see SM and Fig. \ref{fig:S5}).}
The spatial distributions of hardening zones and softening zones are presented in Fig. \ref{fig:5} (c) and (d),respectively.
Most of plastic rearrangements (the purple atoms shown in  Figs. \ref{fig:5}(c) and (d)) accumulated by the AMC are located in the polarized zones, which is insensitive to the $D_\text{min}^{2}$ for selecting plastic rearrangements and the threshold $\Delta \tau_y$ for distinguishing polarized zones (see Fig. \ref{fig:S5} (b) and the corresponding discussion in the SM). 
The strong correlation between polarized zones and local plastic rearrangements confirms that the mechanical polarity induced by AMC is originated from a few local plastic rearrangements and the LYS can be regarded as the structural indicator of the polarized sample.
Moreover, it is clear that the hardening zones and softening zones are almost identical, which means that the rearranged atoms brings distinct influences to the positive and negative directions.
\begin{figure}[!htpb]
\centering
\includegraphics[width=0.8\textwidth]{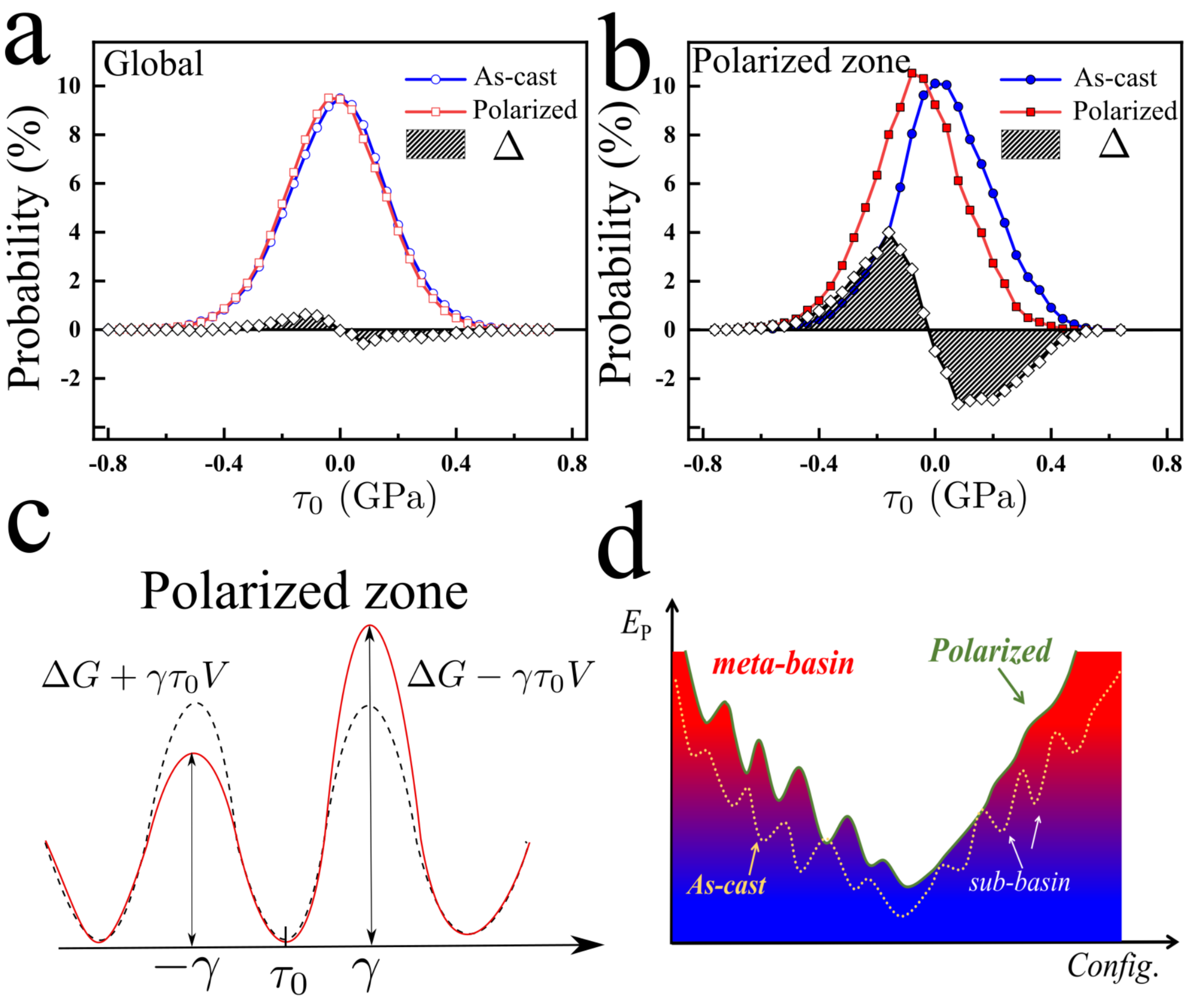}
\caption{{Local residual stress and potential energy surface.
(a) The total distribution of local residual stresses for the polarized and as-cast samples. 
(b) The distributions of local residual stresses in the polarized zones for the polarized and as-cast samples. 
$\Delta$ represent the differences between the polarized and as-cast samples.
(c) Schematic illustration of the local residual stress effects on the potential energy surface of polarized zones.
(d) Schematic illustration of the inherent features of PEL for the as-cast (the dash line) and polarized (the solid line) samples.
}
}
\label{fig:6}
\end{figure}
\section{Discussion and conclusion}

The mechanical anisotropy of BMGs is achieved by the quasi-elastic AMC protocol.
Unlike the previous methods\cite{peng2013stress,PhysRevLett.124.205503} based on the huge plastic deformation, the AMC trained sample is polarized by the local plastic rearrangements rather than a global structural evolution.
Mechanical loading induced polarity is a widespread phenomenon in crystal or amorphous solids, which is termed as Bauschinger effect\cite{weinmann1988bauschinger,PhysRevLett.124.205503}.
Unlike the polarity in crystal solid which is dominated by topological defects such as dislocation or grain boundary, the polarity in metallic glasses is attributed to the topological bound switch \cite{PhysRevB.48.3048} or shear transform zone (STZ)\cite{PhysRevE.57.7192}. In the amorphous solids, the switch events or STZ would release stresses to surrounds \cite{PhysRevLett.103.036001}, and then the stress distribution of two states will be different after plastic rearrangement.
It should emerge the shift of the distribution of local residual stresses in the polarized sample respect to the as-cast sample. 
However, as shown in Fig. \ref{fig:6}(a), the distributions of local residual stresses for the polarized and as-cast samples are almost equivalent.
The averaged local residual stress is around zero which means the polarized sample could be mechanical stable under ambient pressure.
No shift of the distribution indicates that the quasi-elastic AMC protocol did not introduce macroscopic residual stresses into the sample as plastic flow or creep deformation always did.

However, the distribution of local residual stresses in the polarized zones (Fig. \ref{fig:6} (b)) shows that the polarized sample shifts to the left with respect to the as-cast sample.
According to the Eyring model\cite{eyring1936viscosity}, the energy barrier can be influenced by applied stress \cite{PhysRevE.73.061106} and a simple schematic diagram can be proposed. 
The dash line represents the local Frenkel potential surface \cite{frenkel1926theorie,PhysRevLett.95.195501} of one typical polarized zone in the as-cast sample.
The negative local residual stresses increases the energy barrier in positive direction and simultaneously decreases the negative one.
{As a result, though there is no spatial correlation between positive and negative LYS (Figs. \ref{fig:S4}(c),(d)) of the as-cast sample, the softening zones and hardening zones located in two opposite directions have strong spatial correlations as shown in Figs. \ref{fig:5} (c) and (d).}

{The quasi-elastic AMC polarized the sample by the healing of positive STZs associated with the production of negative STZs.
It can be illustratively discribed by the reduced local potential energy surfaces, as shown in Fig. \ref{fig:6}(d).
According to the PEL conception, the amorphous system is trapped in a meta-basin consisting of a number of sub-basins. Since the quasi-elastic AMC only slightly rejuvenates the MG samples, the uplift at the bottom of the meta-basin is not obvious.
For the as-cast sample with isotropic mechanical behaviors, the density of sub-basins is statistically identical in all directions (the dash line in Fig. \ref{fig:6}(d)). 
However, for the polarized sample prepared by quasi-elastic AMC, the morphology of PEL becomes quasi-elastic and sub-basin-free in one direction, but more rugged in the opposite direction (the solid line).
}
{The anelastic arrangements still can be triggered due to the strain-created sub-basins in the quasi-elastic direction, which is consistent with previous works by Xu et al \cite{xu2017strain,PhysRevLett.120.125503} based on the PEL calculation.
The degree of polarity can be well controlled by the normalized  difference between the densities of plastic STZs for two opposite strains.
Thus, the higher density of initial plastic STZs, such as in ultra-rejuvenated or low dimensional MGs, provide a wider range of mechanical regulation.}

{In conclusion, we proposed an quasi-elastic AMC method and used it to achieve mechanical anisotropy of MGs without destroying samples or inducing mechanical annealing or rejuvenation effects.
The degree of polarity and the anelastic limit can be well controlled by regulating the amplitude of mechanical cycling.
We find that only plastic STZs corresponding to the loading direction can be healed by asymmetric cycling and the achieved anisotropic behavior originates from the exhaustion of plastic STZs in the loading direction.
The anelastic rearrangements are still however observable in trained glasses during unloading.
While usual structural indicators are not sensitive probe for induced anisotropy, the asymmetric local yield stresses distribution works well to capture the polarization. 
Moreover, the increase of elastic limit is always accompanied by the decrease of potential energy, when the MG sample is processed by thermal or mechanical annealing.
However, the AMC protocol significantly improves the elastic limit in aimed direction while slightly rejuvenation the mechanical property and other physical properties that are sensitive to energy state, which provides a potential way to achieve  multi-performance regulation of MGs. Our finding is of fundamental importance, which further our understanding of the mechanical deformation of metallic glasses and sheds some light on the prospects for improved properties through induced anisotropy.
}

\section{acknowledgments}
We thank Jean-Louis Barrat, Kirsten Martens, Zheng Wang, En Ma for useful discussion. And we also appreciate Hai-Long Peng to share the code of anisotropy part of RDF with us. This work is supported by Guangdong Major Project of Basic and Applied Basic Research, China (Grant No. 2019B030302010),  the NSF of China (Grant Nos.51601009, {Nos.52130108}, Nos. U1930402),  B.S.S and P.F.G acknowledges the computational support from the Beijing Computational Science Research Center (CSRC).

\pagebreak
\section*{Supplementary Materials}
\title{Supplementary Materials-Cycle deformation enabled controllable mechanical polarity and elastic limit of metallic glasses}
\author{Baoshuang Shang}
\affiliation{Songshan Lake Materials Laboratory, Dongguan 523808, China}
\affiliation{Beijing Computational Science research center, Beijing 100094, China}
\author{Weihua Wang}
\affiliation{Institute of Physics, Chinese Academy of Sciences, Beijing 100190, China}
\affiliation{Songshan Lake Materials Laboratory, Dongguan 523808, China}
\author{Pengfei Guan}
\email{pguan@csrc.ac.cn}
\affiliation{Beijing Computational Science research center, Beijing 100094, China}
\maketitle
\renewcommand{\thefigure}{S\arabic{figure}}
\renewcommand{\theequation}{S\arabic{equation}}
\renewcommand{\thesection}{S\arabic{section}}
\setcounter{figure}{0}
\setcounter{equation}{0}
\setcounter{section}{0}
\section{The local yield stress by frozen matrix method}
\begin{figure}
\centering
\includegraphics[width=0.6\textwidth]{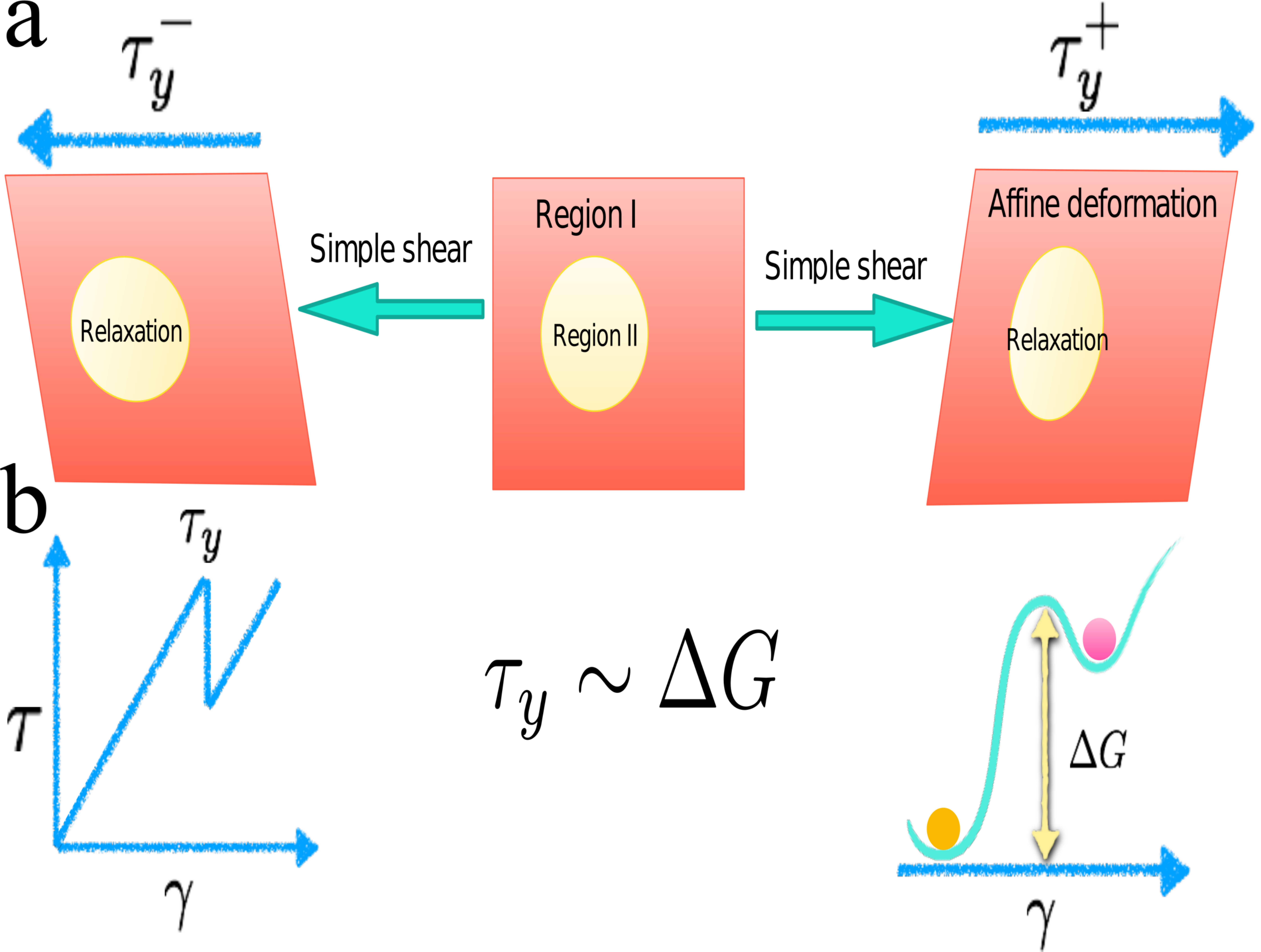}
\caption{
The procedure to calculate LYS. (a) region II is selected to relax under macro deformation and region I is constrained to affine deformation. And there are two directional LYS noted as $\tau_{y}^{-}$ and $\tau_{y}^{+}$, respectively. (b) illustration of the correlation between LYS and local energy barrier, when local stress of region II increase to a critical value, there would be a stress drop, which indicates a energy transit from a local minimum to another, the LYS correlates with energy barrier.
}
\label{fig:S1}
\end{figure}

Fig. \ref{fig:S1}(a) illustrates the process of frozen matrix method. 
The local region II was  selected as investigation zone, and the sample went with one step affine deformation ($\Delta \gamma=10^{-4}$), then only region II was relaxed to satisfy the mechanical equilibrium condition, the process was repeated until the first stress drop appearing, and we recorded the stress value one strain step before first stress drop. 
The first drop in the local stress-strain curve can represent the level of energy barrier (Fig. \ref{fig:S1} (b)).
\section{The mechanical anisotropy and hysteresis loop.}

\begin{figure}[!]
\centering
\includegraphics[width=0.6\textwidth]{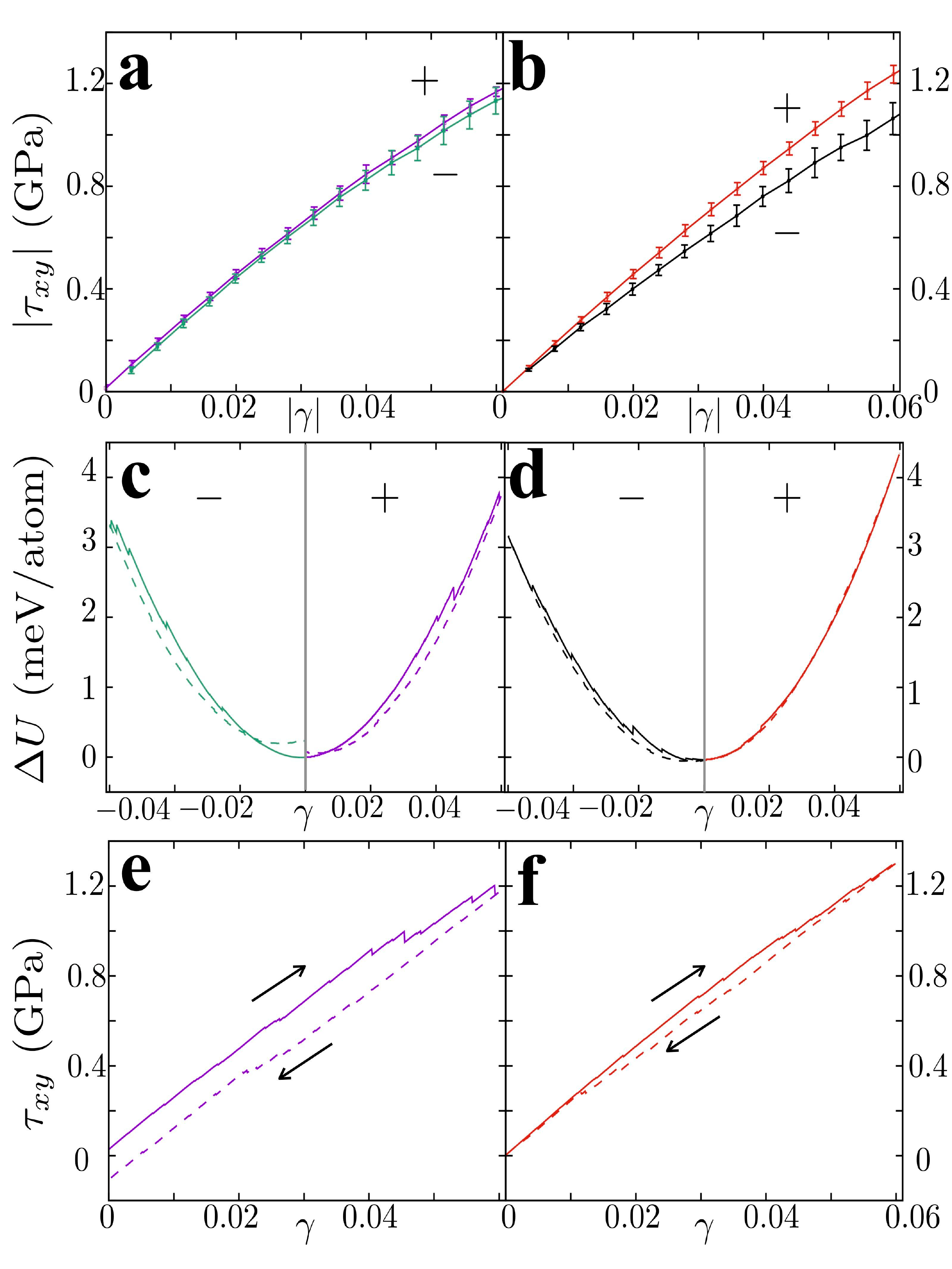}
\caption{Mechanical property of as-cast and polarized samples ($\gamma_{max}=0.06$). (a), (b) the elastic stress-strain curve of as-cast and polarized samples in the two directions : positive ($+$) and negative ($-$). The curve is statistically averaged from ten independent samples, and the errorbar is the standard deviation of all samples. (c), (d) the potential energy evaluates with the strain in positive and negative directions for representative as-cast and polarized sample, respectively. The solid line is the loading direction and the dash line is the unloading direction. (e), (f) the corresponding loading curve along positive direction for (c), (d), the dash line is the unloading curve. The arrow is guided for the eye to show the loading-unloading direction.} 
\label{fig:S2}
\end{figure}

Fig. \ref{fig:S2} shows the mechanical property of as-cast and polarized samples, the as-cast sample is isotropic and polarized sample manifests a notable mechanical anisotropy in stress strain curve. In contrast mechanical anisotropy aroused AMC, the potential energy of polarized sample is slightly changed, and the hysteresis loop is reduced, but the sample shape can be recovered thoroughly in the training direction.

\section{The structual indicators of polarization structure.}
\begin{figure}[!]
  \centering
  \includegraphics[width=0.6\textwidth]{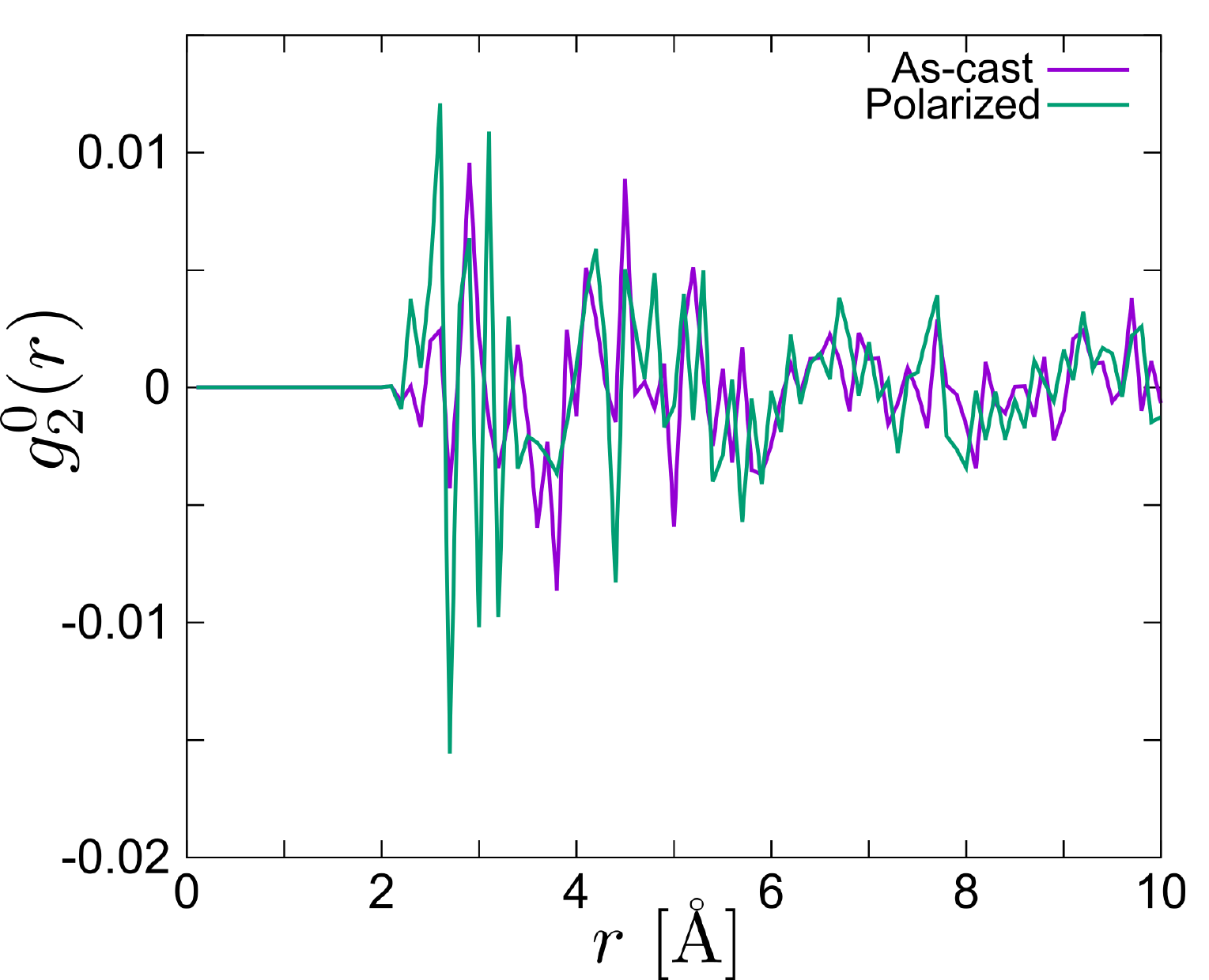}
  \caption{ The anisotropy component of RDF
$g^0_2(r)$, for as-cast and polarized sample, respectively.
}
\label{fig:S3}
\end{figure}

\begin{figure}[!]
\centering
\includegraphics[width=0.6\textwidth]{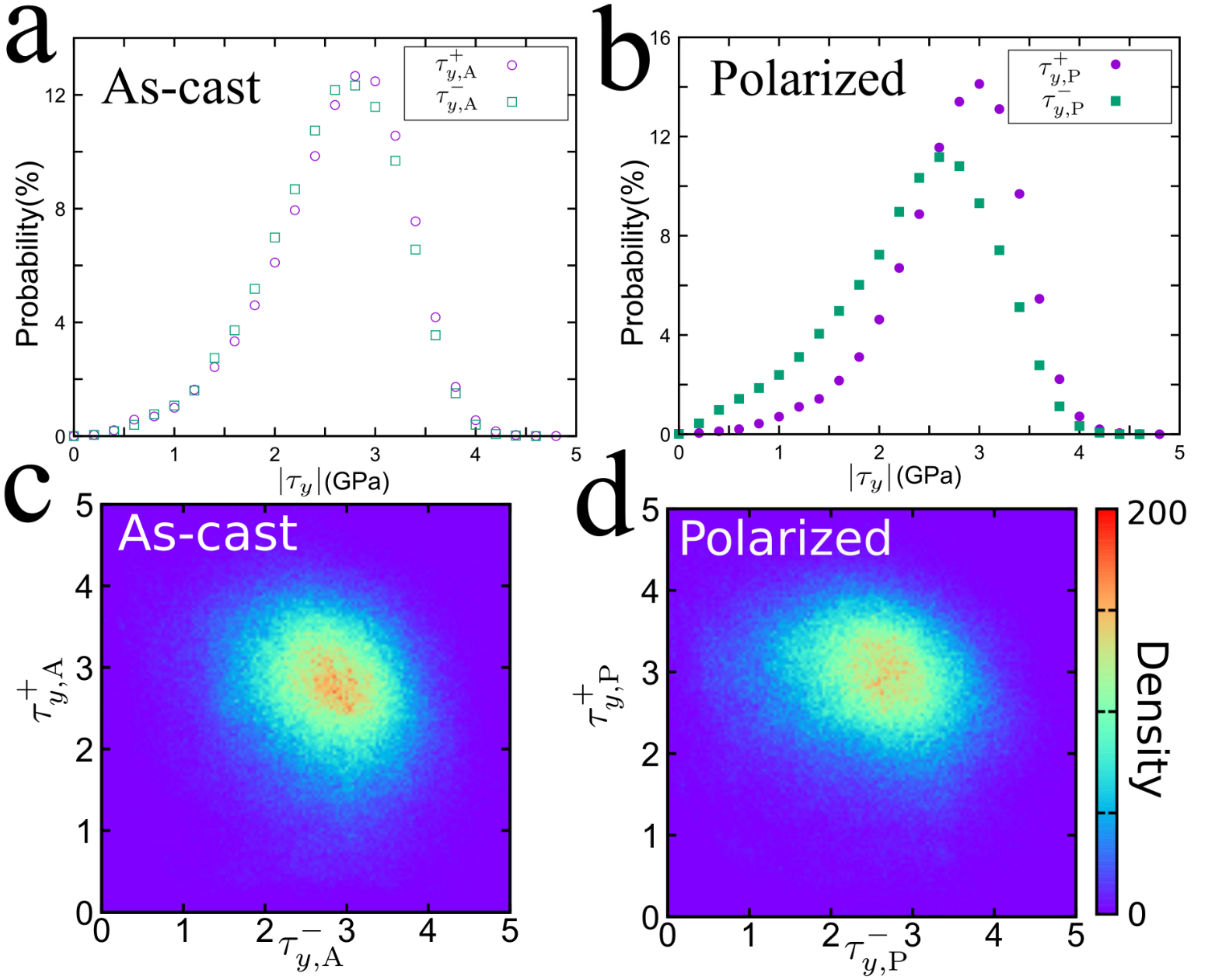}
\caption{The correlation and distribution of local yield stresses. (a),(b) Probability distribution of LYS for as-cast and polarized sample, respectively. (c),(d) the spatial correlation between LYS in negative and positive direction for as-cast and polarized samples, respectively.}
\label{fig:S4}
\end{figure}

The probability distribution of LYS is directional insensitive in the as-cast sample (Fig. \ref{fig:S4}(a)), and in contrast the probability distribution of LYS for the polarized sample has been significantly skewed by the AMC training (Fig. \ref{fig:S4}(b)). 
There is no spatial correlation in LYS between the positive and negative directions, both for the as-cast and polarized samples (Fig. \ref{fig:S4}(c),(d)). 

\section{The Effects  of threshold of polarized zone ($\tau_c$) and plastic rearrangement ($\eta$)}
\begin{figure}[!]
\centering
\includegraphics[width=0.6\textwidth]{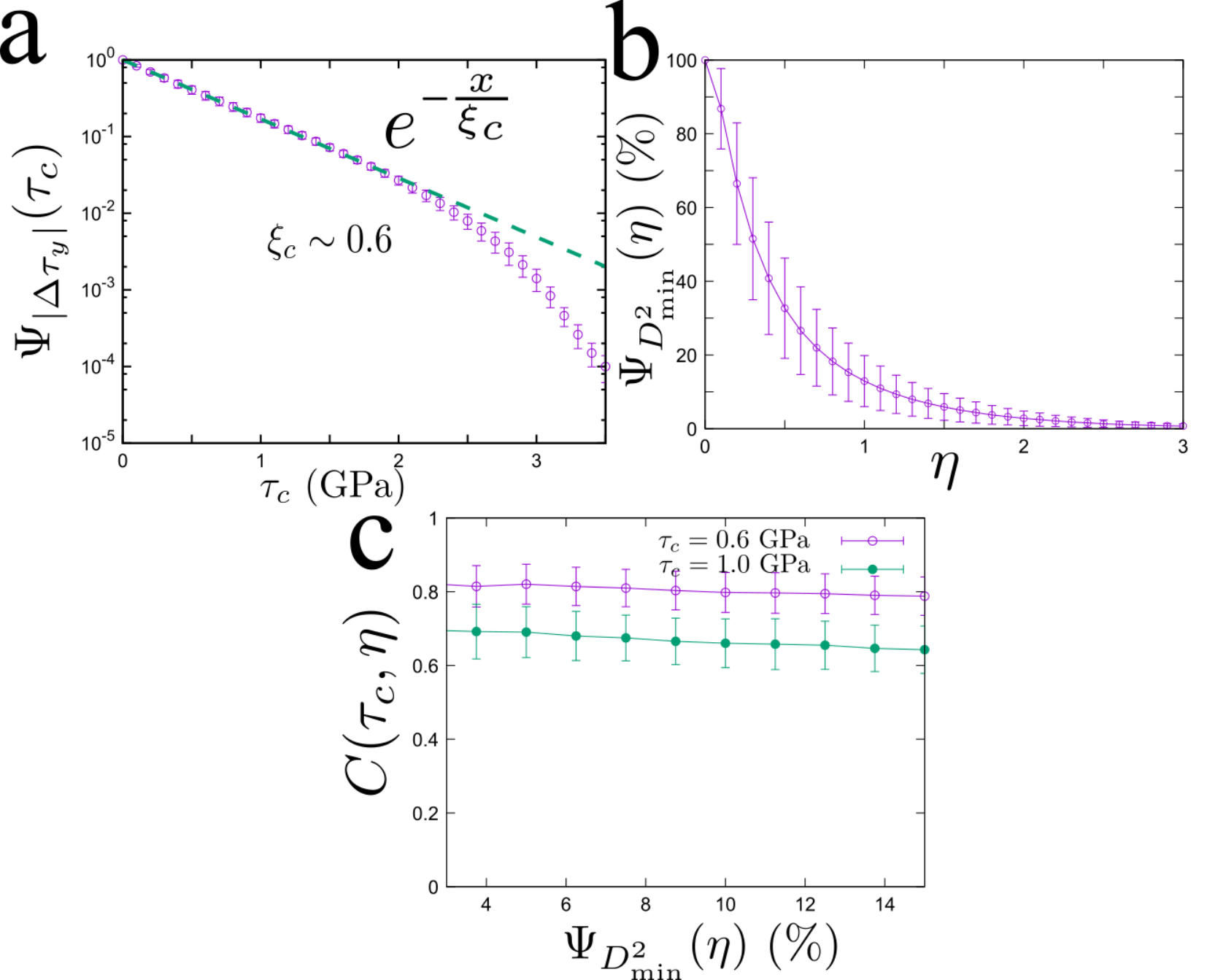}
\caption{The effects of threshold (a). The CCDF of $\Psi_{|\Delta \tau_y|(\tau_c)}$ versus the threshold $\tau_c$. (b). The CCDF of $\Psi_{D_\text{min}^2}(\eta)$ versus the threshold $\eta$. (c). The spatial correlation between polarized zone and plastic rearrangement versus threshold of plastic rearrangement for different threshold of polarized zone.}
\label{fig:S5}
\end{figure}

{The complementary cumulative distribution function (CCDF) $\Psi_X$ is defined as }
\begin{equation*}
  \Psi_x(X)=\int_{X}^{\infty}p(x)dx
\end{equation*}
Where $X$ is the threshold value, and $p(x)$ is the probability density function of variable $x$, it shows the probability of the variable $x$ is larger than $X$.
We calculated the CCDF of deviation between as-cast and polarized LYS ($|\Delta \tau_y| \equiv |\tau_{y,P}-\tau_{y,A}|$, both include the soften region and harden region), $\Psi_{|\Delta \tau_y|}(\tau_c)$, which shows the percentage of the number of atoms involved in the region of $|\Delta \tau_y|>\tau_c$, $\tau_c$ is the threshold value of $|\Delta \tau_y|$. 
It can be well depicted by an exponential decay function ($e^{-x/\xi_c}$) with a cutoff, shown in the dash line in Fig. \ref{fig:S5} (a), and the $\xi_c$ can give a characterize deviation level, the value is about 0.6 GPa.  To emphasize the non-symmetry property in the polarized sample, the characterize distance $\tau_c$ is modified to 1 GPa to exclude some symmetry zones. If the deviation level is larger than the characterize value, it can be recognized as the polarized zone. 

 We also calculated the CCDF of local rearrangement atom $\Psi_{D_\text{min}^2}(\eta)$ between the as-cast and polarized samples (Fig. \ref{fig:S5} (c)), the value $\eta$ is the threshold of plastic rearrangement.  To check the reboust of threthold for polarized zone and plastic rearrangement, we define a correlation function $C(\tau_c,\eta)$ as 
\begin{equation*}
  C(\tau_c,\eta)=\frac{n \in \{\Psi_{D_\text{min}^2}(\eta)\cap \Psi_{|\Delta \tau_y|}(\tau_c)\}}{n \in \{ \Psi_{D_\text{min}^2}(\eta)\}}
\end{equation*}
Where  $n \in \{\Psi_{D_\text{min}^2}(\eta)\}$ is the number of atoms that satisfy the condition $D_\text{min}^2>\eta$, and  $n \in \{\Psi_{D_\text{min}^2}(\eta)\cap \Psi_{|\Delta \tau_y|}(\tau_c)\}$ is the number of atoms that satisfy the conditions $D_\text{min}^2 > \eta$ and $|\Delta \tau_y|>\tau_c$. For $C(\tau_c,\eta)=1$ , it means all the plastic rearrangements happen in the $|\tau_y|>\tau_c$ zone. 
And $C(\tau_c,\eta)=0.0$ means all the plastic rearrangements happen outside the $\Psi_{\Delta \tau_y}(\tau_c)$ zone.
As shown in Fig. \ref{fig:S5}, the correlation between plastic rearrangements and the polarized zones is insensitive with the threshold of corresponding value. Hence in the manuscript, we choose $\Psi_{|D_\text{min}^2|}(\eta)=5\%$ and $\tau_c=1$ GPa
\end{document}